\documentclass[prb,preprint]{revtex4-1} 

\usepackage{amsmath,bm}  
\usepackage{amsfonts} 
\usepackage{graphicx} 
\usepackage{mathrsfs} 
\usepackage{amssymb}
\usepackage{ dsfont }
\usepackage{bbm}
\usepackage[cal=boondox,scr=boondoxo]{mathalfa}
\usepackage[bbgreekl]{mathbbol}
\usepackage{hyperref}

\begin{document}
\preprint{\textsl{KPOP}$\mathscr{E}$-2020-03}

\title{Three-Body Inertia Tensor}

\author{June-Haak Ee}
\email{chodigi@gmail.com}
\affiliation{\textsl{KPOP}$\mathscr{E}$ Collaboration, Department of Physics, Korea University, Seoul 02841, Korea}

\author{Dong-Won Jung}
\email{dongwonj@korea.ac.kr} 
\affiliation{\textsl{KPOP}$\mathscr{E}$ Collaboration, Department of Physics, Korea University, Seoul 02841, Korea}

\author{U-Rae Kim}
\email{sadafada@korea.ac.kr} 
\affiliation{\textsl{KPOP}$\mathscr{E}$ Collaboration, Department of Physics, Korea University, Seoul 02841, Korea}

\author{Dohyun Kim}
\email{tradysori705@korea.ac.kr}
\affiliation{\textsl{KPOP}$\mathscr{E}$ Collaboration, Department of Physics, Korea University, Seoul 02841, Korea}

\author{Jungil Lee}
\email{jungil@korea.ac.kr} 
\thanks{Director of Korea Pragmatist Organization for Physics Education (\textsl{KPOP}$\mathscr{E}$)}
\affiliation{\textsl{KPOP}$\mathscr{E}$ Collaboration, Department of Physics, Korea University, Seoul 02841, Korea}

\date{\today}

\begin{abstract}
We derive a general formula for the inertia tensor of a three-body system.
By employing three independent
Lagrange undetermined multipliers to express the vectors corresponding to
the sides in terms of
the position vectors of the vertices,
we present the general covariant expression for the
inertia tensor of the three particles of different
masses. 
If $m_a/a=m_b/b=m_c/c=\rho$, then
the center of mass coincides with the incenter of the triangle and
the moment of inertia about the normal axis passing the center of mass is
$I=\rho abc$,
where $m_a$, $m_b$, and $m_c$ are the masses of the particles
at $A$, $B$, and $C$, respectively, and 
$a$, $b$, and $c$ are the lengths of the line segments $\overline{BC}$,
$\overline{CA}$, and $\overline{AB}$, respectively.
The derivation and the corresponding results are closely related to the famous 
Heron's formula for the area of a triangle.
\end{abstract}

\maketitle 

\begin{widetext}
\section{Introduction}
The center of mass (CM) and inertia tensor are fundamental physical quantities
for a system of particles. In usual textbooks like Refs.~\cite{CM}
on classical mechanics, two-body systems and specific 
rigid bodies like rods, plates, discs, cubes, and spheres  are considered.
However, students are not frequently exposed to a three-body system that 
constructs a triangle to investigate the fundamental mechanical properties
of the CM and the inertia tensor. 

Normally undergraduate students are well aware of various formulas involving
trigonometric identities that project the fundamental nature of a three-body
system. In this work, we are particularly interested in an arbitrary
three-body system of pointlike masses $m_a$, $m_b$, and $m_c$ placed
at points $A$, $B$, and $C$, respectively. It is straightforward to find
the CM of this three-body system. The computation of the inertia tensor
of the system involves the computation of $m_a\bm{A}^{\star2}+m_b\bm{B}^{\star2}+m_c\bm{C}^{\star2}$,
where  $\bm{A}^\star$, $\bm{B}^\star$, and $\bm{C}^\star$ are the lever-arm vectors
from the CM to the vertices $A$, $B$, and $C$, respectively.
We can express the lever-arm vectors as linear combinations of
vectors $\bm{a}$, $\bm{b}$, and $\bm{c}$ that represent the sides
opposite to the vertices.
Because $\bm{A}^\star$, $\bm{B}^\star$, and $\bm{C}^\star$ are coplanar,
those linear combinations are not unique.
We employ Lagrange's method of undetermined multipliers to find the inverse transformation \cite{Prather-1965,Goedecke-1966,Faddeev-1967,Gaskill-1969,Balart-2019}.
The complicated dependence on the multipliers eventually cancels when we
compute a physically measurable quantities like the scalar products
of any two vectors.

By making use of the bra-ket notation
and completely covariant approach \cite{Ee-2017},
we have computed the inertia tensor in an arbitrary frame of reference.
As a special example, we consider the case in which the CM coincides with the incenter of the triangle. 
In that case, the perpendicular distance from the CM to every side is the same.
We have demonstrated that the condition is satisfied if  $m_a:m_b:m_c=a:b:c$.
The solutions that we discarded involves unphysical mass contributions.
The moment of inertia about the normal axis passing the center of mass is found to be
$I=\rho abc$ when $\rho=m_a/a=m_b/b=m_c/c$. 
The derivation and the corresponding results are closely related to the famous 
Heron's formula for the area of a triangle.

This paper is organized as follows:  In Sec.~\ref{sec:def}, we list the definitions of
various physical variables that we frequently use in the remainder of this paper.
Section \ref{sec:multiplier} contains the computation of the scalar products
of the lever-arm vectors by employing Lagrange's method of undetermined multipliers.
The features of the special case that the CM coincides with the incenter are considered
in Sec.~\ref{sec:incenter}. The relationship among our mechanical variables and
Heron's formula for the area of a triangle is presented in Sec.~\ref{sec:Heron}.
The computation of the inertia tensor of the three-body system is given in Sec.~\ref{sec:IT}
that is followed by the discussion in Sec.~\ref{sec:discussion}.
\section{Definitions\label{sec:def}}
As shown in Fig.~\ref{figure:masses},
there are three particles of mass $m_a$, $m_b$,  and $m_c$
placed at  $\bm{A}$, $\bm{B}$, and $\bm{C}$, respectively. We denote
the total mass by $M=m_a+m_b+m_c$. 
\begin{figure}
\centering
\includegraphics[width=0.8\columnwidth]{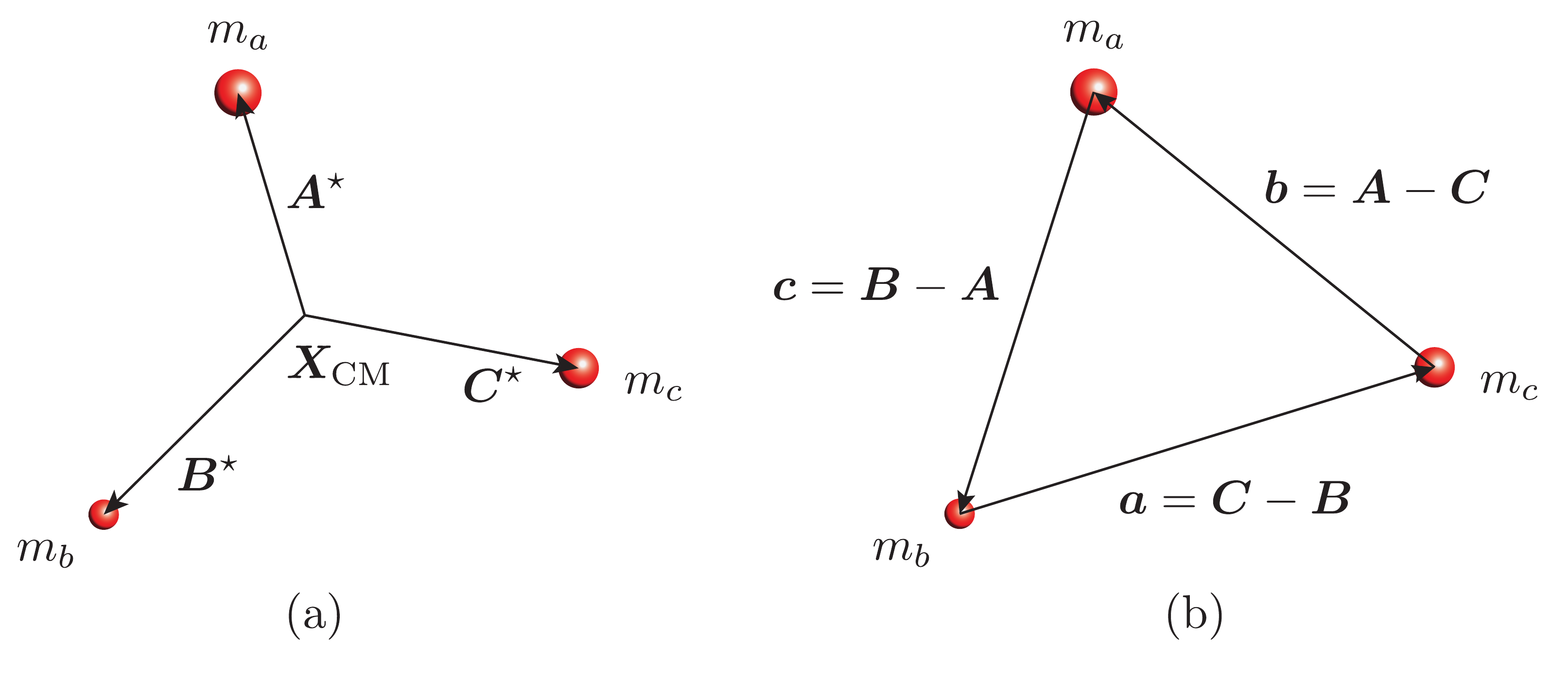}
\caption{\label{figure:masses}
(a)
The lever-arm vectors $\bm{A}^\star$, $\bm{B}^\star$, and $\bm{C}^\star$
from the CM, $\bm{X}_{\rm CM}$, to three particles of mass $m_a$, $m_b$, and $m_c$,
respectively.
(b) The side vectors $\bm{a}$, $\bm{b}$, and $\bm{c}$ defined in Eq.~(\ref{abcABC}),
where $\bm{V}=\bm{X}_{\rm CM}+\bm{V}^{\star}$.}
\end{figure}
The CM \cite{CM} is at
\begin{eqnarray}
\bm{X}_{\rm CM}
=\frac{1}{M}(m_a\bm{A}+m_b\bm{B}+m_c\bm{C}).
\end{eqnarray}
We employ the bra-ket notation \cite{Dirac-1923,Dirac-1939,Dirac-1950,Dirac-1958}
 $\langle \bm{U}|\bm{V}\rangle\equiv \bm{U}\cdot\bm{V}$
for any Euclidean vectors $\bm{U}$ and $\bm{V}$.
The inertia tensor operator can then be defined by
\begin{eqnarray}
\label{TOP}
\mathbf{I}=(m_a\bm{A}^2+m_b\bm{B}^2+m_c\bm{C}^2)\mathbf{1}
-m_a|\bm{A}\rangle\langle\bm{A}|
-m_b|\bm{B}\rangle\langle\bm{B}|
-m_c|\bm{C}\rangle\langle\bm{C}|.
\end{eqnarray}
Here, $\mathbf{1}$ is the identity operator whose matrix representation
is $\mathbbm{1}=(\delta_{ij})$.
In any Cartesian coordinates with the unit basis vectors $\hat{\bm{e}}_i$,
the inertia tensor is the matrix element of $\mathbbm{I}=(I_{ij})$ 
that can be projected out from (\ref{TOP}) as
\begin{equation}
I_{ij}=\langle \hat{\bm{e}}_i|\mathbf{I}|\hat{\bm{e}}_j\rangle
=m_a\left[\delta_{ij}\left(\sum_{k=1}^{3} A_k^2\right)-A_i A_j\right]
+m_b\left[\delta_{ij}\left(\sum_{k=1}^{3} B_k^2\right)-B_i B_j\right]
+m_c\left[\delta_{ij}\left(\sum_{k=1}^{3} C_k^2\right)-C_i C_j\right]
,
\end{equation}
where $\langle \hat{\bm{e}}_i|\hat{\bm{e}}_j\rangle=\delta_{ij}$,
and $V_i= \langle \hat{\bm{e}}_i |\bm{V}\rangle
=\langle \bm{V} | \hat{\bm{e}}_i\rangle$ for any Euclidean vector $\bm{V}$.

We attach a superscript $\star$ to represent a vector in the CM frame such that
\begin{equation}
\bm{V}^{\star}=\bm{V}-\bm{X}_{\rm CM}.
\end{equation}
Thus in the CM frame, the three \textit{lever-arm vectors} $\bm{A}^\star$, $\bm{B}^\star$, and $\bm{C}^\star$
from the CM
are coplanar:
\begin{eqnarray}
\label{XatCM}
m_a\bm{A}^{\star}+m_b\bm{B}^{\star}+m_c\bm{C}^{\star}=\bm{0}.  
\end{eqnarray}

Parallel-axis theorem can be employed to express the inertia tensor as
\begin{eqnarray}
\label{TOP2}
\mathbf{I}&=&(M\bm{X}_{\rm CM}^2+m_a\bm{A}^{\star2}+m_b\bm{B}^{\star2}+m_c\bm{C}^{\star2})\mathbf{1}
\nonumber\\
&&
-M|\bm{X}_{\rm CM}\rangle\langle\bm{X}_{\rm CM}|
-m_a|\bm{A}^{\star}\rangle\langle\bm{A}^{\star}|
-m_b|\bm{B}^{\star}\rangle\langle\bm{B}^{\star}|
-m_c|\bm{C}^{\star}\rangle\langle\bm{C}^{\star}|.
\end{eqnarray}

We define the displacements $\overrightarrow{BC}$,
$\overrightarrow{CA}$, and $\overrightarrow{AB}$ that are independent of the frame
of reference as
\begin{subequations}
\label{abcABC}
\begin{eqnarray}
\bm{a}&=&\overrightarrow{BC}=\bm{C}-\bm{B}=\bm{C}^{\star}-\bm{B}^{\star},
\\
\bm{b}&=&\overrightarrow{CA}=\bm{A}-\bm{C}=\bm{A}^{\star}-\bm{C}^{\star},
\\
\bm{c}&=&\overrightarrow{AB}=\bm{B}-\bm{A}=\bm{B}^{\star}-\bm{A}^{\star}.
\end{eqnarray}
\end{subequations}

\section{Lagrange Multipliers and Invariants\label{sec:multiplier}}
We want to express the lever-arm vectors
$\bm{A}^{\star}$, $\bm{B}^{\star}$, and $\bm{C}^{\star}$ in terms of
the \textit{side vectors}
$\bm{a}$, $\bm{b}$, and $\bm{c}$.
According to \eqref{XatCM}, 
only two of the three
 vectors $\bm{A}^{\star}$, $\bm{B}^{\star}$, and $\bm{C}^{\star}$ are linearly
 independent. Thus the inverse transformation is not unique.
We introduce three Lagrange undetermined multipliers 
$\lambda_1$,  $\lambda_2$, and  
$\lambda_3$ with the dimensions of mass inverse.  
By adding $\lambda_i(m_a\bm{A}^\star+m_b\bm{B}^\star+m_c\bm{C}^\star)$,
which is actually vanishing according to (\ref{XatCM}),
to the constraint equations in Eq.~(\ref{abcABC}),
we can make the transformation invertible:
\begin{equation}
\begin{pmatrix}
\bm{a}\\
\bm{b}\\
\bm{c}
\end{pmatrix}
=\mathbb{\Lambda}(\lambda_1,\lambda_2,\lambda_3)
\begin{pmatrix}
 \bm{A}^{\star}\\
 \bm{B}^{\star}\\
 \bm{C}^{\star}
\end{pmatrix}, 
\end{equation}
where the matrix $\mathbb{\Lambda}(\lambda_1,\lambda_2,\lambda_3)$
is defined by
\begin{equation}
\mathbb{\Lambda}(\lambda_1,\lambda_2,\lambda_3)=
\begin{pmatrix}
0&-1&1\\
1&0&-1\\
-1&1&0 
\end{pmatrix}
+
\begin{pmatrix}
\lambda_1m_a&\lambda_1m_b&\lambda_1m_c\\
\lambda_2m_a&\lambda_2m_b&\lambda_2m_c\\
\lambda_3m_a&\lambda_3m_b&\lambda_3m_c 
\end{pmatrix}.
\end{equation}
While 
$\mathbb{\Lambda}(0,0,0)$ is not invertible,
$\mathbb{\Lambda}(\lambda_1,\lambda_2,\lambda_3)$
has the inverse:
\begin{eqnarray}
\bm{A}^{\star}
&=&
\frac{1}{\mathscr{D}}
\big\{
[1-m_c \lambda_2+m_b \lambda_3]\bm{a}+
[1+m_c \lambda_1+(m_b+m_c) \lambda_3]\bm{b}
+
[1-m_b \lambda_1-(m_b+m_c) \lambda_2]\bm{c}\big\},
\nonumber\\
\bm{B}^{\star}
&=&
\frac{1}{\mathscr{D}}
\big\{
[1-m_c \lambda_2-(m_a+m_c)\lambda_3]\bm{a}+
[1+m_c\lambda_1-m_a\lambda_3]\bm{b}
+
[1+(m_a+m_c)\lambda_1+m_a\lambda_2]\bm{c}\big\},
\nonumber\\
\bm{C}^{\star}
&=&
\frac{1}{\mathscr{D}}
\big\{
[1+(m_a+m_b)\lambda_2+m_b\lambda_3]\bm{a}+
[1-(m_a+m_b)\lambda_1-m_a\lambda_3]\bm{b}
+
[1-m_b\lambda_1+m_a\lambda_2]\bm{c}\big\},
\nonumber\\
\end{eqnarray}
where the determinant $\mathscr{D}$ 
of $\mathbb{\Lambda}(\lambda_1,\lambda_2,\lambda_3)$
is given by
\begin{equation}
\mathscr{D}=
(m_a+m_b+m_c) (\lambda_1+\lambda_2+\lambda_3).
\end{equation}
It is manifest that  the determinant $\mathscr{D}$ 
vanishes when $\lambda_1+\lambda_2+\lambda_3=0.$ Except for
that case, we are free to choose any values $\lambda_i$'s.
The introduction of the multipliers brings in parameter dependence
because the coefficient of each parameter that is actually vanishing
is shared by $\bm{A}^{\star}$, $\bm{B}^{\star}$, and $\bm{C}^{\star}$ nonuniformly.
The explicit dependence on the parameters $\lambda_i$'s
is present only at the vector level.
We shall find that the  parameter  dependence disappears
once we compute the invariant quantities like
scalar products.

According to (\ref{abcABC})  and Fig.~\ref{figure:masses}(b), 
the angle between $\bm{a}$ and $\bm{b}$ is $\pi-\angle C$ and so on.
 Here, we denote the internal angle of the triangle at vertex $\bm{C}$ by $\angle C$, and similarly for $\angle A$ and $\angle B$. 
Therefore, the scalar products of unit vectors 
$\hat{\bm{a}}$, $\hat{\bm{b}}$, and $\hat{\bm{c}}$
parallel to $\bm{a}$, $\bm{b}$, and $\bm{c}$, respectively, are
\begin{subequations}
\label{cosLAW}
\begin{eqnarray}
\hat{\bm{a}}\cdot\hat{\bm{b}}&=&-\cos\angle C= \frac{c^2-a^2-b^2}{2ab},
\\
\hat{\bm{b}}\cdot\hat{\bm{c}}&=&-\cos\angle A= \frac{a^2-b^2-c^2}{2bc},
\\
\hat{\bm{c}}\cdot\hat{\bm{a}}&=&-\cos\angle B= \frac{b^2-c^2-a^2}{2ca}.
\end{eqnarray}
\end{subequations}
While each of the lever-arm vectors  $\bm{A}^{\star}$,  $\bm{B}^{\star}$, and 
 $\bm{C}^{\star}$  are dependent on the parameters $\lambda_i$,
 the scalar products are free of the parameter dependence:
\begin{eqnarray}
\bm{A}^{\star 2}&=&
\frac{m_b+m_c}{D}
\big(    m_b c^2 + m_c b^2  -  \mu_{bc} a^2
\big) , 
\nonumber\\
\bm{B}^{\star 2}&=&
\frac{m_c+m_a}{D}
\big(  m_c a^2 + m_a c^2  -  \mu_{ca} b^2
\big) , 
\nonumber\\
\bm{C}^{\star 2}&=&
\frac{m_a+m_b}{D}
\big(  m_a b^2 + m_b a^2   -  \mu_{ab} c^2
\big) , 
\nonumber\\
\bm{A}^\star\cdot\bm{B}^\star&=&
\frac{m_c}{2D}
\left[
(m_a-m_b+m_c) a^2 +  (m_b-m_a+m_c) b^2  - \left(    m_a+m_b+m_c +\frac{2 m_a m_b}{m_c} \right)  c^2
\right],
\nonumber\\
\bm{B}^\star\cdot\bm{C}^\star&=&
\frac{m_a}{2D}
\left[
(m_b-m_c+m_a) b^2 +  (m_c-m_b+m_a) c^2  - \left(   m_a+m_b+m_c +\frac{2 m_b m_c}{m_a} \right)  a^2
\right],
\nonumber\\
\bm{C}^\star\cdot\bm{A}^\star&=&
\frac{m_b}{2D}
\left[ 
(m_c-m_a+m_b) c^2 +  (m_{a}-m_{c}+m_b) a^2  - \left(    m_a+m_b+m_c +\frac{2 m_c m_a}{m_b} \right)  b^2
\right],
\nonumber\\
\label{SCALARS}
\end{eqnarray}
where $D=(m_a+m_b+m_c)^2$
and $\mu_{ij}=m_im_j/(m_i+m_j)$ 
is the reduced mass of a two-body system.

\section{When CM is identical to the Incenter\label{sec:incenter}}
We consider a special case in which the CM becomes identical to the incenter of the triangle.
We shall find that
the CM is identical to the incenter of the triangle at which
the following ratios are all equal:
\begin{equation}
\label{rhoabc}
\rho=\frac{m_a}{a}=\frac{m_b}{b}=\frac{m_c}{c}.
\end{equation}

Let us introduce 
 the position vector $\bm{A}'$ for the foot of
the normal from the $\bm{X}_{\rm CM}$ to $\bm{a}$. $\bm{B}'$ and $\bm{C}'$
are defined in a similar manner as shown in Fig.~\ref{figure:normal}.
\begin{figure}
\centering
\includegraphics[width=0.4\columnwidth]{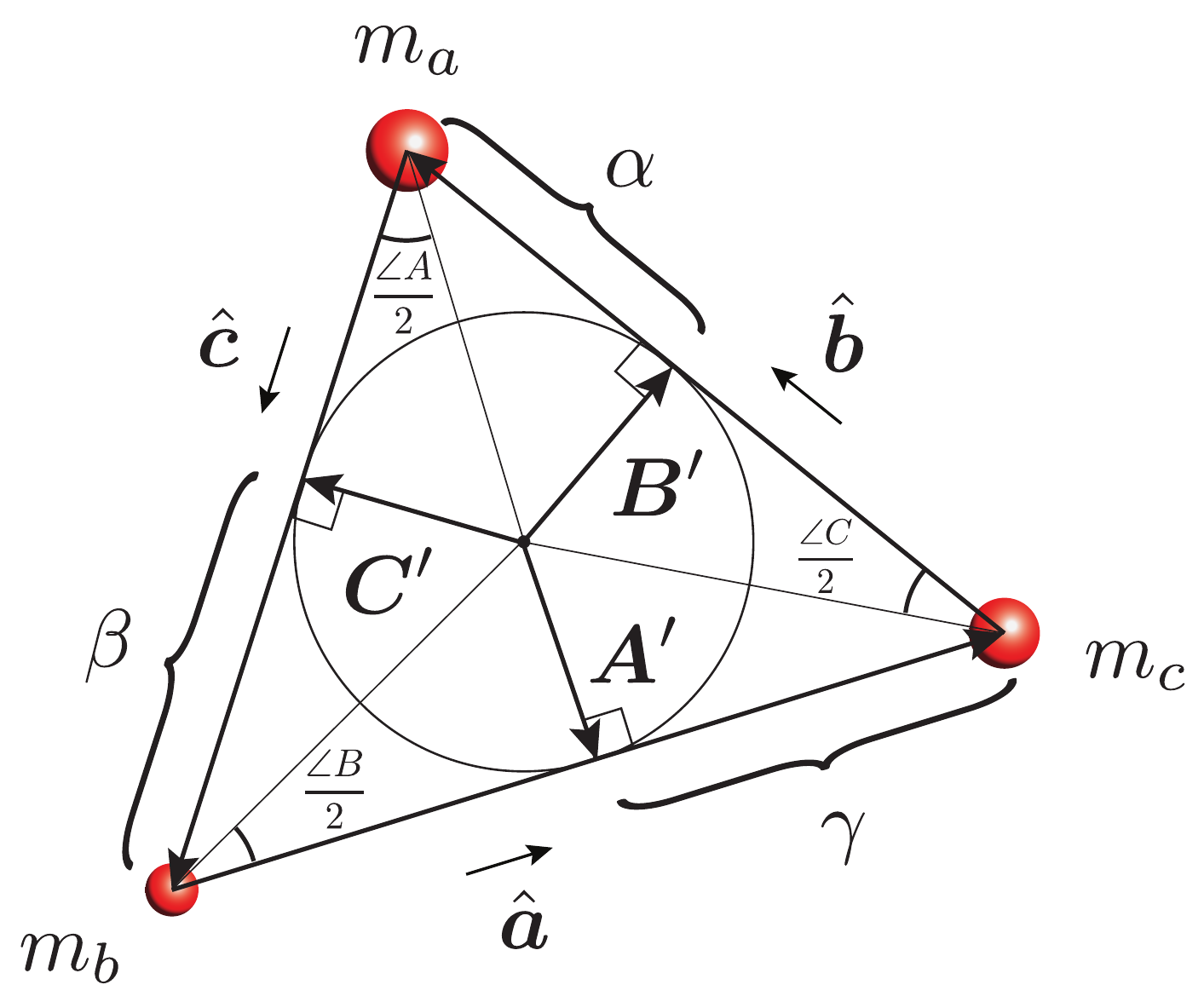}
\caption{\label{figure:normal}
The definitions of normal vectors $\bm{A}'$, $\bm{B}'$, and $\bm{C}'$.
We show the directions of the unit vectors 
$\hat{\bm{a}}$, $\hat{\bm{b}}$, and $\hat{\bm{c}}$,
and the lengths $\alpha$, $\beta$, and $\gamma$ defined in Eq.~(\ref{eq:define-albegam}).
We also show the half angles at the vertices $A$, $B$, and $C$.
}
\end{figure}
Then the vectors $\bm{A}^\star$, $\bm{B}^\star$, and $\bm{C}^\star$ bisect the angles $\angle A$,
$\angle B$, and $\angle C$, respectively.
We define 
\begin{eqnarray}
\label{eq:define-albegam}
\alpha&=&|\overrightarrow{AB'}|=|\overrightarrow{AC'}|,
\\
\beta&=&|\overrightarrow{BC'}|=|\overrightarrow{BA'}|,
\\
\gamma&=&|\overrightarrow{CA'}|=|\overrightarrow{CB'}|.
\end{eqnarray}
Then the \textit{normal vectors}
$\bm{A}'$ ,
$\bm{B}'$ , and
$\bm{C}'$  are expressed as
\begin{eqnarray}
\bm{A}'&=&\bm{B}^\star+\beta \hat{\bm{a}}=\bm{C}^\star-\gamma \hat{\bm{a}},
\\
\bm{B}'&=&\bm{C}^\star+\gamma \hat{\bm{b}}=\bm{A}^\star-\alpha \hat{\bm{b}},
\\
\bm{C}'&=&\bm{A}^\star+\alpha \hat{\bm{c}}=\bm{B}^\star-\beta \hat{\bm{c}}.
\end{eqnarray}
Hence,
\begin{eqnarray}
\bm{a} &=&a\hat{\bm{a}}=(\beta+\gamma)\hat{\bm{a}},
\\
\bm{b} &=&b\hat{\bm{b}}=(\gamma+\alpha)\hat{\bm{b}},
\\
\bm{c} &=&c\hat{\bm{c}}=(\alpha+\beta)\hat{\bm{c}}.
\end{eqnarray}

We define the perpendicular distance from the CM to $\bm{a}$, $\bm{b}$, and $\bm{c}$ by
\begin{subequations}
\label{rabc}
\begin{eqnarray}
r_a 
&=&\sqrt{\bm{A}^{\prime 2}}
=\sqrt{\bm{B}^{\star2}}\sin\frac{\angle B}{2}=\beta\tan\frac{\angle B}{2},
 \\
r_b &=&\sqrt{\bm{B}^{\prime 2}}
=\sqrt{\bm{C}^{\star2}}\sin\frac{\angle C}{2}=\gamma\tan\frac{\angle C}{2},
 \\
r_c &=&\sqrt{\bm{C}^{\prime 2}}
=\sqrt{\bm{A}^{\star2}}\sin\frac{\angle A}{2}=\alpha\tan\frac{\angle A}{2}.
\end{eqnarray}
\end{subequations}
By making use of the trigonometric identity $\sin\frac{\theta}{2} 
= \sqrt{(1-\cos\theta)/2}$, we find $r_a^2$, $r_b^2$, and $r_c^2$
from Eqs.~(\ref{cosLAW}) and (\ref{SCALARS}) as follows:
\begin{subequations}
\label{r2abc}
\begin{eqnarray}
r_a^2&=&
Fb(b+c-a) (b-c+a)
(m_c+m_a)
\big(    m_c a^2 + m_a c^2    -  \mu_{ca} b^2
\big),
\\
r_b^2&=&
Fc(c+a-b) (c-a+b)
(m_a+m_b)
\big(    m_a b^2 + m_b a^2   -  \mu_{ab} c^2
\big),
\\
r_c^2&=&
Fa (a+b-c) (a-b+c)
(m_b+m_c)
\big(    m_b c^2 + m_c b^2   -  \mu_{bc} a^2
\big),
\end{eqnarray}
\end{subequations}
where $F\equiv1/[4abc(m_a+m_b+m_c)^2]$.

If the position of the CM coincides with 
the incenter of the triangle,  then  $r_a=r_b=r_c$.
The first solution is given by
\begin{equation}
\label{mabc-ans}
m_a:m_b:m_c=a:b:c.
\end{equation}
The other solutions are
\begin{subequations}
\begin{eqnarray}
\label{mimaginary}
m_a:m_b:m_c
&=&
b^2-a^2-c^2 \pm\sqrt{\lambda(a^2,b^2,c^2)}:
a^2-b^2-c^2 \mp\sqrt{\lambda(a^2,b^2,c^2)}:
2c^2,
\\
\label{mimaginary2}
m_a:m_b:m_c
&=&
a^2 (b^2+c^2-a^2-b c):
 b^2 (a^2+c^2-b^2-a c): c^2(b^2+a^2-c^2-a b ),
\phantom{X}
\end{eqnarray}
\end{subequations}
where K\"all\'en function
\cite{Dalitz-1954,Dalitz-1958,Kallen-1964,Venturi-1974}
$\lambda(a^2,b^2,c^2)$ is defined by
\begin{eqnarray}
\label{lambda-def}
\lambda(a^2,b^2,c^2)
&=& -(a+b+c)(a+b-c)(b+c-a)(c+a-b)
\nonumber\\
&=&  a^4+b^4+c^4-2(a^2b^2+b^2c^2+c^2a^2).
\end{eqnarray}

According to the first line of (\ref{lambda-def}),
$\lambda(a^2,b^2,c^2)$ must be negative definite because
$a+b+c>0$,
$a+b>c$,
$b+c>a$, and
$c+a>b$.
Thus we discard the solution (\ref{mimaginary}) which gives imaginary mass.
We also discard the solution of Eq.~(\ref{mimaginary2}) that involves
a negative mass.
Listing the sides in order as $c\ge b \ge a$,
we can show that
\begin{eqnarray}
b^2+c^2-a^2-b c
&\le &b^2+c^2-a^2-b^2=c^2-a^2,
\nonumber \\
a^2+c^2-b^2-a c
&\le &a^2+c^2-b^2-a^2=c^2-b^2,
\nonumber \\
b^2+a^2-c^2-a b
&\le &b^2+a^2-c^2-a^2 = b^2-c^2\le 0.
\label{imcondi}
\end{eqnarray}
By imposing the conditions in \eqref{imcondi} 
into Eq.~(\ref{mimaginary2}), 
we find that
at least one particle must be massless or of negative mass.
Thus, we discard this unphysical solution, too.
As a result, (\ref{mabc-ans}) is the unique physical solution for $r_a^2=r_b^2=r_c^2$
and is consistent with (\ref{rhoabc}). 
We substitute the constraint (\ref{rhoabc}) into (\ref{r2abc}) to find that
$r_a=r_b=r_c=r$, where
\begin{eqnarray}\label{rabcf}
r&\equiv&
\sqrt{ \frac{(a+b-c) (b+c-a) (c+a-b)}{4 (a+b+c) }} 
=\sqrt{\bm{A}^{\prime2}}=\sqrt{\bm{B}^{\prime2}}=\sqrt{\bm{C}^{\prime2}}.
\end{eqnarray}

According to (\ref{rabc}), we find that
\begin{subequations}
\label{abcs}
\begin{eqnarray}
\alpha&=&\frac{r}{\tan\frac{\angle A}{2}}
=\frac{1}{2}(b+c-a)=s-a, 
 \\
\beta&=&\frac{r}{\tan\frac{\angle B}{2}}
=\frac{1}{2}(c+a-b)=s-b, 
 \\
\gamma&=&\frac{r}{\tan\frac{\angle C}{2}}
=\frac{1}{2}(a+b-c)=s-c, 
\end{eqnarray}
where we used the trigonometric identity
$\tan\frac{\theta}{2} = \sqrt{(1-\cos\theta)/(1+\cos\theta)}$
and define $s$ as
\begin{eqnarray}
s=\alpha+\beta+\gamma=\frac{1}{2}(a+b+c).
\end{eqnarray}
\end{subequations}
We substitute the constraint (\ref{rhoabc}) into
the scalar products in Eq.~(\ref{SCALARS}) to 
determine the scalar products of the lever-arm vectors:
\begin{subequations}
\label{SCALARSabc}
\begin{eqnarray}
\bm{A}^{\star 2}&=&
\frac{b c (b+c-a)}{ a+b+c },\phantom{X}
\\
\bm{B}^{\star 2}&=&
\frac{c a (c+a-b)}{ a+b+c },
\\
\bm{C}^{\star 2}&=&
\frac{a b (a+b-c)}{ a+b+c },
\\
\bm{A}^\star\cdot\bm{B}^\star&=&
-\frac{c (c+a-b)(c-a+b)}{2(a+b+c) },
\\
\bm{B}^\star\cdot\bm{C}^\star&=&
-\frac{a (a+b-c)(a-b+c)}{2(a+b+c) },
\\
\bm{C}^\star\cdot\bm{A}^\star&=&
-\frac{b (b+c-a)(b-c+a)}{2(a+b+c) }.
\end{eqnarray}
\end{subequations}

\section{Heron's Formula\label{sec:Heron}}
By making use of the identities in Eq.~(\ref{cosLAW}), we  obtain
\begin{subequations}
\begin{eqnarray}
\sin \angle A 
&=&\sqrt{1-\cos^2\angle A}=
\frac{\sqrt{-\lambda(a^2,b^2,c^2)}}{2bc},
\\
\sin \angle B 
&=&\sqrt{1-\cos^2\angle B}=
\frac{\sqrt{-\lambda(a^2,b^2,c^2)}}{2ca},
\\
\sin \angle C 
&=&\sqrt{1-\cos^2\angle C}=
\frac{\sqrt{-\lambda(a^2,b^2,c^2)}}{2ab},
\end{eqnarray}
\end{subequations}
where K\"all\'en function
$\lambda(a^2,b^2,c^2)$ is defined in Eq.~(\ref{lambda-def}).
Heron's formula for the area of a triangle is reproduced:
\begin{eqnarray}
\Delta&=&\frac{1}{2}ab\sin \angle C
=\frac{1}{2}bc\sin \angle A
=\frac{1}{2}ca\sin \angle B
\nonumber\\
&=&\frac{1}{4}
\sqrt{-\lambda(a^2,b^2,c^2)}
\nonumber\\
&=&\frac{1}{4}\sqrt{(a+b+c)(a+b-c)(b+c-a)(c+a-b)}
\nonumber\\
&=&rs
\nonumber\\
&=&\sqrt{s(s-a)(s-b)(s-c)} 
\nonumber\\
&=&\frac{1}{2}r(a+b+c)
\nonumber\\
&=&r(\alpha+\beta+\gamma),
\end{eqnarray}
where 
we have made use of the identities (\ref{rabcf}) and (\ref{abcs}).

\section{Inertia Tensor\label{sec:IT}}
The inertia tensor for arbitrary masses and length parameters can be
computed by making use of (\ref{TOP2}). In this section, we compute
that tensor.
One of the principal axis is along 
$\hat{\bm{n}}$ perpendicular to the triangle passing
the CM. 
The moment of inertia about that axis 
is
\begin{eqnarray}
I_{\hat{\bm{n}}}=
m_a \bm{A}^{\star2}+m_b \bm{B}^{\star2}+m_c \bm{C}^{\star2}.
\end{eqnarray}
Because $\bm{A}^{\star}$, $\bm{B}^{\star}$, and $\bm{C}^{\star}$ are coplanar and $\hat{\bm{n}}$ is perpendicular
to the triangle, the coefficient of $|\hat{\bm{n}}\rangle\langle\hat{\bm{n}}|$ must
be the same as that for $\mathbf{1}$:
\begin{eqnarray}
\mathbf{I}
&=&I_{\hat{\bm{n}}}(|\hat{\bm{n}}\rangle\langle\hat{\bm{n}}|+\mathbf{1}_\perp)
- m_a|{\bm{A}^\star}\rangle\langle{\bm{A}^\star}|
- m_b|{\bm{B}^\star}\rangle\langle{\bm{B}^\star}|
- m_c|{\bm{C}^\star}\rangle\langle{\bm{C}^\star}|,
\end{eqnarray}
where $\mathbf{1}_\perp$ is the two-dimensional 
identity operator on the triangle plane.
We choose the cyclic orthonormal basis vectors as
\begin{eqnarray}
\hat{\bm{e}}_1 &=& \frac{\bm{A}^\star}{\sqrt{\bm{A}^{\star2}}},
\nonumber \\
\hat{\bm{e}}_2 &=&\hat{\bm{n}}\times \hat{\bm{e}}_1.
\end{eqnarray}
Then $\mathbf{1}_\perp=
|\hat{\bm{e}}_1\rangle \langle\hat{\bm{e}}_1|
+|\hat{\bm{e}}_2\rangle \langle\hat{\bm{e}}_2|$
and
\begin{eqnarray}
\mathbf{I}
&=&I_{\hat{\bm{n}}} |\hat{\bm{n}}\rangle\langle\hat{\bm{n}}|
+\sum_{i,\,j=1}^2
I_{ij} |\hat{\bm{e}}_i\rangle \langle\hat{\bm{e}}_j|,
\end{eqnarray}
where 
\begin{eqnarray}
I_{11} &=& m_b B_\perp^{2}+m_c C_\perp^{2},
\nonumber \\
I_{12} &=& -m_b B_\parallel B_\perp - m_c C_\parallel
C_\perp
=
I_{21},
\nonumber \\
I_{22} &=& m_a A_\parallel^{2}+m_b B_\parallel^{2} +m_c C_\parallel^{2}.
\end{eqnarray}
Here, $X_\parallel$ and $X_\perp$  are the components of 
$\bm{X}^\star$ 
that are parallel and perpendicular to $\bm{A}^\star$, respectively,
as shown in Fig.~\ref{figure:tensor}. 
\begin{figure}
\centering
\includegraphics[width=0.4\columnwidth]{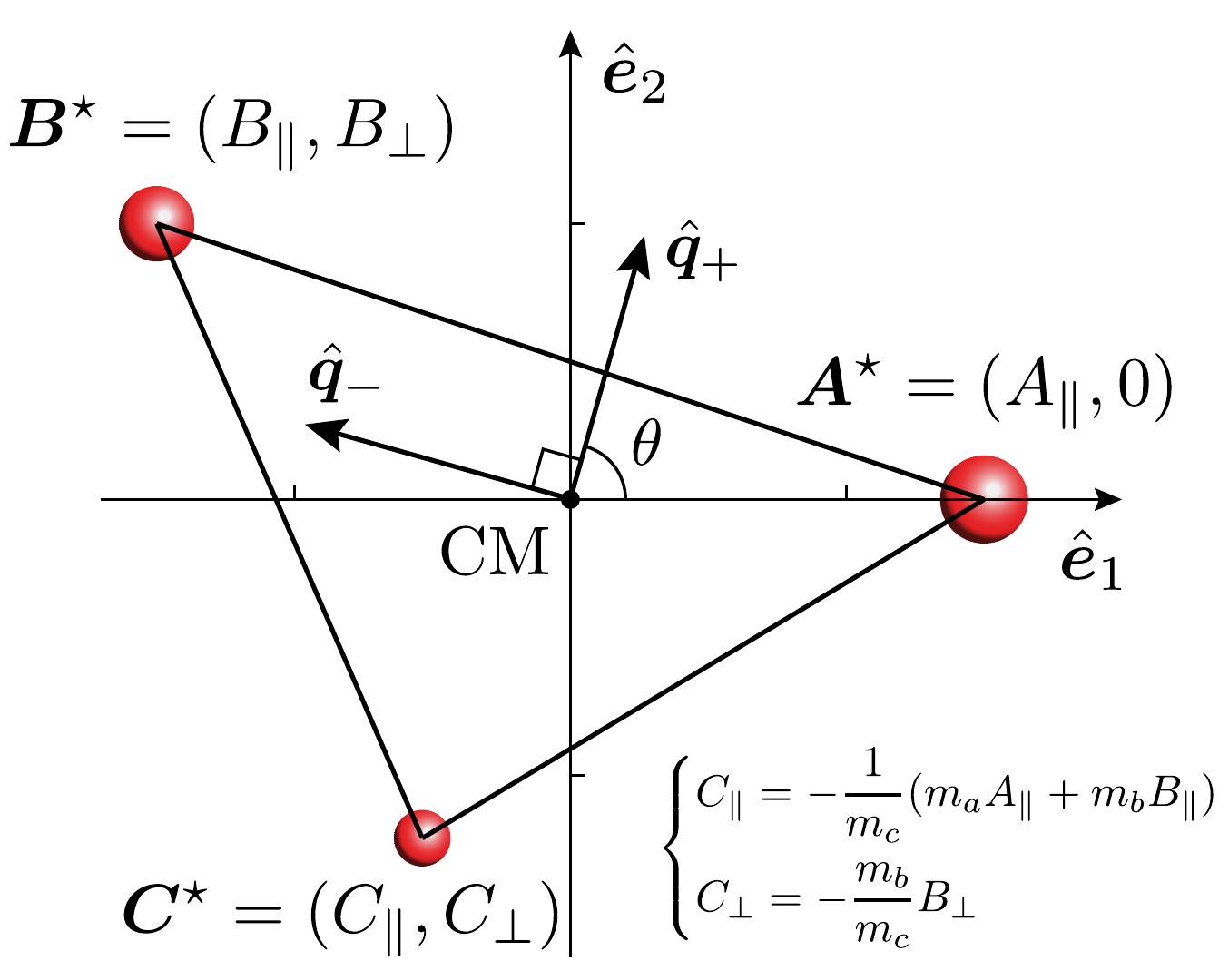}
\caption{\label{figure:tensor}
The lever-arm vectors of the three particles in the frame 
where $\bm{A}^\star$ is along the $x$ axis.
The principal axes $\hat{\bm{q}}_+$ and $\hat{\bm{q}}_-$ 
can be obtained
by rotating $\hat{\bm{e}}_1$ and $\hat{\bm{e}}_2$, respectively, by an angle $\theta$ 
defined in \eqref{eq:theta-def}.}
\end{figure}
We can replace the components of $\bm{C}^\star$ with those of 
$\bm{A}^\star$ and $\bm{B}^\star$ by making use of the CM condition
$\bm{C}^{\star}=-(m_a\bm{A}^{\star}+m_b\bm{B}^{\star})/m_c$.

The inertia tensor operator becomes most compact
if we choose the principal axes to be the basis vectors
$\hat{\bm{q}}_\pm$ where
\begin{eqnarray}
\mathbf{I}
&=&I_{\hat{\bm{n}}} |\hat{\bm{n}}\rangle\langle\hat{\bm{n}}|
+I_+ |\hat{\bm{q}}_+\rangle\langle\hat{\bm{q}}_+|
+I_- |\hat{\bm{q}}_-\rangle\langle\hat{\bm{q}}_-|.
\end{eqnarray}
The principal axes $\{\hat{\bm{q}}_+,\hat{\bm{q}}_-\}$
can be obtained by rotating $\{\hat{\bm{e}}_1,\hat{\bm{e}}_2\}$
as
\begin{eqnarray}
\label{principal12}
|\hat{\bm{q}}_+\rangle
&=&\phantom{+}\cos\theta|\hat{\bm{e}}_1\rangle 
+ \sin\theta|\hat{\bm{e}}_2\rangle,
\nonumber \\
|\hat{\bm{q}}_-\rangle
&=&-\sin\theta|\hat{\bm{e}}_1\rangle +\cos\theta|\hat{\bm{e}}_2\rangle,
\end{eqnarray}
where the angle $\theta$ is 
\begin{eqnarray}
\label{eq:theta-def}
\theta = \frac{1}{2}\arctan\frac{2I_{12}}{I_{11}-I_{22}}.
\end{eqnarray}
The principal moments are now determined as
\begin{eqnarray}
\label{master}
I_{\hat{\bm{n}}}
&=&
m_a \bm{A}^{\star2}+m_b \bm{B}^{\star2}+m_c \bm{C}^{\star2},
\nonumber
\\
I_+
&=&
I_{11} \cos^2\theta + I_{22} \sin^2\theta + I_{12} \sin2\theta,
\nonumber \\
I_-
&=&
I_{22} \cos^2\theta + I_{11} \sin^2\theta - I_{12} \sin2\theta.
\end{eqnarray}

The rotationally invariant forms of 
$\bm{A}^\star$, $\bm{B}^\star$, and $\bm{C}^\star$ 
are
\begin{subequations}
\label{eq:ABC-comp}
\begin{eqnarray}
A_\parallel &=& \sqrt{\bm{A}^{\star2}},
 \\
B_\parallel &=& \frac{\bm{A}^\star \cdot \bm{B}^\star}{\sqrt{\bm{A}^{\star2}}},
 \\
B_\perp &=&
\frac{(\bm{A}^\star \times \bm{B}^\star)\cdot \hat{\bm{n}}}
{\sqrt{\bm{A}^{\star2}}},
 \\
C_\parallel &=& \frac{\bm{A}^\star \cdot \bm{C}^\star}{\sqrt{\bm{A}^{\star2}}},
 \\
C_\perp &=&
\frac{(\bm{A}^\star \times \bm{C}^\star)\cdot \hat{\bm{n}}}
{\sqrt{\bm{A}^{\star2}}}.
\end{eqnarray}
\end{subequations}
Substituting Eq.~(\ref{eq:ABC-comp}) into Eq.~(\ref{master})
and applying Eq.~(\ref{SCALARS}),
we express the principal moments $I_{\hat{\bm{n}}}$,
$I_+$, and $I_-$ in terms of only the masses and lengths:
\begin{subequations}
\begin{eqnarray}
\label{eq:In-result}
I_{\hat{\bm{n}}}&=&
\frac{m_a m_b c^2+m_b m_c a^2 + m_c m_a b^2}{m_a+m_b+m_c},
 \\
\label{eq:Ipm-result}
I_\pm 
&=&
\frac{I_{\hat{\bm{n}}}}{2}
\pm
\frac{1}{2}\sqrt{I_{\hat{\bm{n}}}^2
+\frac{m_am_bm_c}{m_a+m_b+m_c}\lambda(a^2,b^2,c^2)},
\end{eqnarray}
\end{subequations}
where the K\"all\'en function $\lambda(a^2,b^2,c^2)$ 
is defined in Eq.~(\ref{lambda-def}).
It is manifest that
the above expressions
satisfy
the perpendicular-axis theorem,
$I_++I_-=I_{\hat{\bm{n}}}$.

If the incenter of the triangle coincides with the CM,
then the condition Eq.~(\ref{rhoabc}) 
requires
\begin{eqnarray}
\label{Iij7}
I_{\hat{\bm{n}}} &=& \rho a b c,
\nonumber \\
I_\pm
&=&
\frac{\rho abc}{2}
\pm
\frac{\rho abc}{2}\sqrt{1
-\frac{(a+b-c)(b+c-a)(c+a-b)}{abc}}.
\end{eqnarray}

\section{Discussion\label{sec:discussion}}
We have investigated the fundamental properties of a three-body system by
computing the inertia tensor in a completely covariant way. 
The calculation is greatly simplified by employing the bra-ket notation.
The inertia tensor is expressed in terms of the masses $m_a$, $m_b$, $m_c$ and 
vectors $\bm{a}$, $\bm{b}$, $\bm{c}$ representing the sides opposite to
the three particles, respectively. The lever-arm vectors $\bm{A}^\star$, $\bm{B}^\star$, and $\bm{C}^\star$
that represent the displacements from the CM to particles are expressed in terms
of $\bm{a}$, $\bm{b}$, and $\bm{c}$ by introducing three independent Lagrange undetermined
multipliers $\lambda_1$, $\lambda_2$, and $\lambda_3$.
The principal axes  consist of the normal vector $\hat{\bm{n}}$
and the two placed on the triangle plane: $\bm{q}_+$ and $\bm{q}_-$ shown in Eq.~(\ref{principal12}).
The general expressions for the 
principal moment corresponding to the normal vector $\hat{\bm{n}}$
in Eq.~(\ref{eq:In-result})
is proportional to the weighted sum of the side lengths squared 
with the weighting factor $1/m_i$.
The expressions for the 
principal moments on the triangle plane in Eq.~(\ref{eq:Ipm-result})
involve the K\"all\'en function $\lambda(a^2,b^2,c^2)$
which is negative definite 
for the side lengths $a$, $b$, and $c$.

Lagrange's method of undetermined multipliers is frequently used to determine constraint forces
in Lagrangian mechanics. It has been applied to the quantum field theory for 
a gauge field that has the propagator of vanishing determinant. \cite{Faddeev-1967}
The gauge-fixing term being added to the Lagrangian density is indeed the
application of Lagrange's method of undetermined multiplier. 
The gauge invariance of
the theory guarantees that the final gauge-invariant expression must be free of
the multiplier while the intermediate gauge-dependent expressions like the Feynman rules
may contain the multiplier.
It is particularly useful to solve a system of linear equations
that has multiple solutions. While the introduction of the multipliers brings in
complicated intermediate results at the vector level, our final results for the
inertia tensor are free of these parameters because the results are expressed in terms
of the scalar products that are invariant. Our employment of the undetermined
multiplier should be a heuristic example with which students can be familiar
with the method.

Our derivation for the inertia tensor involved a full exploitation of
Dirac's bra-ket notation. The inertia tensor operator was expressed as the
sum of operators before the projection onto a specific coordinate system.
Thus during the full procedure of the derivation, we were free of complicated 
vector indices. Extension to the fully covariant derivation \cite{Ee-2017} can be directly
obtained from the operator representation. Explicit coordinate dependence
can then be projected out from the covariant expression.
Ordinarily, undergraduate physics major students are exposed
to the bra-ket notation for the first time when they take a course on
quantum mechanics. We argue that a lot of experiences can be made regarding
the bra-ket notation when they study classical mechanics.

As a specific example, we have considered the case in which the CM coincides with the incenter 
of the triangle. In that case, the perpendicular distance from the CM to every side
is the same.
We have demonstrated that the condition is satisfied if 
$m_a:m_b:m_c=a:b:c$.
The derivation and the corresponding results are closely related to the famous 
Heron's formula for the area of a triangle. 
While Heron's formula is for the area of
a triangle, the formula $I_{\hat{\bm{n}}}=\rho abc$
in Eq.~(\ref{Iij7})
is actually a derivative of Heron's formula extended to the moment of inertia.

\begin{acknowledgments}
As members of the Korea Pragmatist Organization for Physics Education
(\textsl{KPOP}$\mathscr{E}$), 
the authors thank to the remaining members of \textsl{KPOP}$\mathscr{E}$ 
for useful discussions.
This work is supported in part by the National Research Foundation
of Korea (NRF) grant funded by the Korea government (MSIT) 
under Contract Nos.\,NRF-2020R1A2C3009918 (J.-H.E., U-R.K., D.W.J., D.K., 
and J.L.),
NRF-2017R1E1A1A01074699 (D.W.J. and J.L.), 
NRF-2018R1D1A1B07047812 (D.W.J.), and  NRF-2019R1A6A3A01096460 (U-R.K.).
\end{acknowledgments}

\end{widetext}

\end{document}